\catcode`\@=11
{\count255=\time\divide\count255 by 60
\xdef\hourmin{\number\count255}
\multiply\count255 by-60\advance\count255 by\time
\xdef\hourmin{\hourmin:\ifnum\count255<10 0\fi\the\count255}}
\def\ps@draft{\let\@mkboth\@gobbletwo \def\@oddhead{} \def\@oddfoot
{\hbox to 7 cm{$\scriptstyle Draft\ version:\ \draftdate$ \hfil} \hskip
-7cm\hfil\rm\thepage \hfil}
\def\@evenhead{}\let\@evenfoot\@oddfoot} \catcode`\@=12
\def\draftdate{\number\month/\number\day/\number\year\ \ \ \hourmin }

\documentstyle[12pt]{article}
\newcommand{\BE}{\begin{eqnarray}}
\newcommand{\EN}{\end{eqnarray}}
\newcommand{\be}{\begin{equation}}
\newcommand{\en}{\end{equation}}

\newcommand{\no}{\noindent}
\newcommand{\vs}{\vspace}

\newcommand{\Bbb}{\bf}
\newcommand{\ha}{{1\over 2}}

\input mssymb.tex

\begin{document}
\title{Stable Non-Gaussian Diffusive Profiles}

\author{J.Bricmont\thanks
{Supported by EC grant SC1-CT91-0695}\\UCL, Physique Th\'eorique,
Louvain-la-Neuve, Belgium\and
A.Kupiainen\thanks{Supported by NSF grant DMS-8903041} \\Helsinki
University, Mathematics Department,\\ Helsinki, Finland}

\date{}

\maketitle
\begin{abstract}
We prove two stability results for the
scale invariant solutions
of the nonlinear heat equation $\partial_t u=\Delta u - |u|^{p-1}u$ with
$1<p<1+{2\over n}$, $n$ being the spatial dimension. The first result is
that a small perturbation of a scale invariant solution vanishes as
$t\rightarrow\infty$. The second result is global, with a positivity
condition on the initial data. \end{abstract}

\section{Introduction}
\setcounter{equation}{0}
%\draft

In \cite{BKL}, we investigated
the long-time asymptotics of
the solutions of nonlinear heat
equations of the type
\be
\partial_t u = \Delta u + F (u, \nabla u, \nabla\nabla u).
\label{2}
\en
and proved that, for a large class of nonlinearities $F$, the solution
behaved, as $t\rightarrow\infty$,
\be
u(x,t)\sim At^{-{1\over2}}f^*({x\over\sqrt{t}}) \label{1}
\en
with
$f^*$ Gaussian, $f^*(x)=e^{-{x^2\over 4}}$. The asymptotics was thus proven
to be ``universal'', i.e. independent on the initial data $u(x,0)$ and the
nonlinearity $F$, within some class. The only dependence of them occurs in
the constant $A$.
We explained this universality,
following the work of
Barenblatt \cite{Ba} and of Goldenfeld et al \cite{Go1,Go2}, in terms of
the Renormalization Group, which also was an important ingredient in the
proof (see Section 3).

The assumptions for the data in \cite{BKL} was smallness in a suitable norm
implying falloff at infinity (for a similar approach with initial data not
decaying at infinity, i.e. the formation of fronts and patterns, see
\cite{bk1,bk2} and for large data blowing up in a finite time, see
\cite{bk3}). The assumptions for $F$ excluded terms such as $u^p$, for
$1<p<1+{2\over n}$,
where $n$ is the dimension of space ($p=1+{2\over n}$ was considered and in
this case the constant $A$ in (\ref{1}) is replaced by $A({\log
t})^{-\ha}$).

Here we want to consider the asymptotics of (\ref{2}) in the presence of
these ``relevant'' (in the Renormalization Group terminology)
terms in $F$. Although again rather general $F$ may be considered, we
formulate below our results for the nonlinear heat equation
\be
\partial_t u = \Delta u - |u|^{p-1} u
\label{4}
\en
where $u=u(x,t), x \in \Bbb R^n$, $1<p<1+{2\over n}$, and we discuss
generalizations in Section 3.

Equation (\ref{4}) has a one-parameter family of scale invariant solutions
(see Section 2) \be
u(x,t) = t^{- \frac{1}{p-1}} f_\gamma (xt^{- \frac{1}{2}}) \label{3}
\en
where $\gamma\geq\gamma_p>0$, provided $1<p<1+{2\over n}$ and $f_\gamma$ is
non-Gaussian, having the asymptotics \be
f_{\gamma}(x) \sim |x|^{- \frac{2}{p-1}} \label{7}
\en
as $|x|\rightarrow\infty$ if
$\gamma > \gamma_p$,
while, for $\gamma= \gamma_p$, it
decays at infinity
as
\be
f_{\gamma_p}(x) \sim |x|^{\frac{2}{p-1}-n} e^{- \frac{x^2}{4}}.
\label{8}
\en

We prove here that these
solutions are stable in two senses: first, there exists a ball in a Banach
space
of initial data such that the
corresponding solutions tend, in the appropriate norm, as
$t\rightarrow\infty$, to (\ref{3}).
Secondly, any initial data satisfying a suitable positivity condition will
give rise to a solution again tending to (\ref{3}).

More precisely, let $q>{2\over p-1}$ and consider the Banach space $B$ of
$L^\infty$ functions $h$ equipped with the norm (with some abuse of
notation!) \be
\| h \|_\infty = {\rm ess}\sup_\xi |h(\xi) (1+|\xi|^q)|. \label{19}
\en
We consider the initial data (taken at time 1 for later convenience)
\be
u(x,1)=f_\gamma(x)+h(x)
\label{89}
\en
with $h\in B$. We prove the

\vs{3mm}

\no {\bf Theorem } {\it Let $1 < p < 1 + \frac{2}{n} $. There exist
$\varepsilon
> 0, C < \infty$ and $\mu > 0$ such that,
if the initial
data $u(x,1)$ of}
(\ref{4}) {\it is given by} (\ref{89}) {\it with $h\in B$ and satisfies
either
$$
\|h\|_\infty \leq \varepsilon
$$
or
$$
h(x)\geq 0
$$
(a.e.)
then,} (\ref{4}) {\it has a unique classical solution and, for all $t$,
$$
\| t^{\frac{1}{p-1}}
u(\cdot t^{\frac{1}{2}},t) - f_\gamma(\cdot) \|_\infty \leq C t^{- \mu}\|
h\|_\infty
$$}

\vs{3mm}

\section{Proof}

\medskip

Before going to the proof of the Theorem, we will briefly discuss the scale
invariant solutions (\ref{3}). These are given by $f_\gamma(x) =
\phi_\gamma(|x|)$ and $\phi_\gamma$ solves the
ordinary differential equation
\be
\phi^{''} + \left( \frac{n-1}{\eta} +
\frac{\eta}{2} \right) \phi^{'} +
\frac{\phi}{p-1} - \phi^{p} = 0
\label{6}
\en
for $\eta = |x| \in [0, \infty[$.

The theory of positive solutions of (\ref{6}) has been developped in
\cite{Br,Ga,KP1}. The main result is that, for any $p>1$, there exists
smooth, everywhere positive solutions, $\phi_\gamma$, of (\ref{6}) with
$\phi_\gamma^{'}(0)=0$ and $\phi_\gamma(0) =\gamma$ for $\gamma$ larger
than
a certain critical
value $\gamma_p$ (but not too large). Actually, for $p < 1 + \frac{2}{n},
\gamma_p > 0$
while $\gamma_p = 0$ for $p \geq 1 + \frac{2}{n}$. The decay at infinity of
these solutions is given in (\ref{7}, \ref{8}).

The existence of a critical $\gamma_p$ can be understood intuitively by
viewing (\ref{6}) as Newton's equation for a particle of mass one, whose
``position" as a function of ``time" is $\phi(\eta)$. The potential
is then $U(\phi) = \frac{\phi^{2}}{2(p-1)} - \frac{\phi^{p+1}}{p+1}$ and
the ``friction term" $\left( \frac{n-1}{\eta} + \frac{\eta}{2} \right)
\phi^{'}$ depends on the ``time" $\eta$. Hence, if $\phi_\gamma^{'}
(0) = 0$ and $\phi_\gamma(0)=\gamma$ is large enough, the time it takes to
approach zero is long and, by then, the friction term has become
sufficiently strong to prevent ``overshooting". However, as $p$ increases,
the potential becomes flatter and one therefore expects $\gamma_p$
to decrease with $p$.

Given the initial data (\ref{89}), it is convenient to rewrite (\ref{4})
in terms of the variables $\xi = xt^{- \frac{1}{2}}$ and $\tau = \log t$;
so,
define $v(\xi, \tau)$ by:
\be
u(x,t) = t^{- \frac{1}{p-1}}
(f_\gamma (xt^{- \frac{1}{2}}) + v (xt^{- \frac{1}{2}}, \log t))
\label{90}
\en
where now
\be
v(\xi,0) = h(\xi).
\label{10}
\en
Then, (\ref{4}) is equivalent to the equation \be
\partial_\tau v = {\cal L} v - \left(|f_\gamma+v|^{p-1}(f_\gamma +v) -
f_\gamma^p -
pf_\gamma^{p-1} v \right) \equiv {\cal L} v + N(v) \label{123}
\en
where we used the fact that (\ref{3}) solves (\ref{4}) and gathered the
linear terms in
$$
{\cal L} = {\cal L}_0 + V_\gamma,
$$
with
\be
{\cal L}_0=\Delta + \frac{\xi}{2} \cdot
\nabla + \frac{1}{p-1},
\label{11}
\en
and
\be
V_\gamma(\xi)=-pf_\gamma^{p-1}(\xi).
\label{124}
\en

To prove that the solution $t^{- \frac{1}{p-1}} (f_\gamma (xt^{-
\frac{1}{2}}) )$ is stable means to find a class of initial data $v(\xi,0)$
such that the corresponding
solution of (\ref{123}) goes to zero as $\tau \rightarrow \infty$, in a
suitable norm.

The Theorem of Section 1 reads now in terms of $v$ as

\vs{3mm}

\no{\bf Proposition 1.} {\it With the assumptions of the Theorem,}
(\ref{123}), {\it with initial data (11) ,has a unique classical solution
and}
$$
\|v(\cdot,\tau)\|_\infty\leq Ce^{-\mu\tau}\|h\|_\infty $$

\vs{3mm}

The main input in the proof is the following estimate on the semigroup
$e^{\tau{\cal L}}$:

\vs{3mm}

\no{\bf Proposition 2.} {\it The operator $e^{\tau{\cal L}}$ is
a bounded operator
in the Banach space $ B$, and its norm satisfies $$
\|e^{\tau{\cal L}} \|\leq Ce^{-\mu\tau}
$$
for some $\mu>0$, $C<\infty$.}

\vs{3mm}

There are two important ingredients in the proof of Proposition 2. The
first is the fact that $e^{\tau{\cal L}}$ is a contraction in a suitable
Hilbert space of rapidly decreasing functions. To see this, note first that
${\cal L}_0$ is conjugate to the Schr\"odinger operator
\be
e^{\frac{\xi^2}{8}} {\cal L}_0 e^{- \frac{\xi^2}{8}} = \Delta -
\frac{\xi^2}{16} -
\frac{n}{4} + \frac{1}{p-1}
\label{12}
\en
i.e. the harmonic oscillator. Thus ${\cal L}_0$ is self-adjoint
on its domain
${\cal D}({\cal L}_0) \subset L^2 ({\bf
R}^n, d\mu)$, where
$$
d\mu(\xi)=e^{\frac{\xi^2}{4}}d\xi.
$$
${\cal L}_0$ has a pure point spectrum $\{{1\over p-1}-{n\over 2} -{m\over
2}\;|\; m=0,1,\dots\}$ and the largest eigenvalue $ {1\over p-1}-{n\over
2}$ is {\it positive} if $1<p<1+{2\over n}$. Thus $e^{\tau{\cal L}_0}$ is
{\it not} contractive and, for $e^{\tau({\cal L}_0+V_\gamma)}$ to contract,
we need to use the potential in a
non-trivial way (this is the reason why $1<p<1+{2\over n}$ is harder than
the $p>1+{2\over n}$ case).

Remarkably, it is possible to prove that ${\cal L}< -E <0$ without a
detailed study of the function $f_\gamma$, but only using equation
(\ref{6}). We have the

\vs{3mm}

\no{\bf Lemma 1.} {\it The operator $e^{\tau{\cal L}}$ is a bounded
operator in the Hilbert space $L^2 ({\bf R}^n, d\mu)$ and its norm
satisfies
$$
\|e^{\tau{\cal L}} \|\leq e^{-E\tau}
$$
for some $E>0$.}

\vs{3mm}

\no{\bf Proof.} Since $V_\gamma$ is bounded, $\cal L$ is self-adjoint and,
as for ${\cal L}_0$, its resolvent is compact and, therefore, its spectrum
is
pure point. Let $-E_\gamma$ be the largest eigenvalue.

First note that $-E_\gamma\leq -E_{\gamma_p}$. Indeed, this holds since
$V_\gamma\leq V_{\gamma_p}$, because
$f_{\gamma}\geq f_{\gamma_p}$, which in turn follows from the fact that
$\gamma_p$ is the smallest
allowed value of $\phi_\gamma(0)=\gamma$ in (\ref{6}), and that two
solutions of (\ref{6}), both with initial conditions $\phi_\gamma'(0)=0$,
will not cross.
Hence it suffices to prove the claim for $\gamma=\gamma_p$. Let us write $E
\equiv E_{\gamma_p}$.

Next we note that, by the Feynman-Kac formula \cite{Si}, $e^{\tau{\cal L}}$
has a strictly positive kernel; indeed, since $-C<V_\gamma(\xi)<0$, we have
\be
e^{-\tau C}e^{\tau{{\cal L}_0}}(\xi,\xi')\leq e^{\tau{\cal
L}}(\xi,\xi')\leq e^{\tau{{\cal L}_0}}(\xi,\xi') \label{fk}
\en
and $e^{\tau{{\cal L}_0}}(\xi,\xi')$ is explicit, see (\ref{69}) below.
Thus, by the
Perron-Frobenius
theorem \cite{GJ}, ${\cal L}$ has a unique eigenvector $\Omega$ with
eigenvalue $-E$
and $\Omega$ can be chosen to be
strictly positive.
That $-E$ is strictly negative, is now shown by the following argument.

Assume for a moment that
$f\equiv f_{\gamma_p} \in {\cal D} ({\cal L})$, and write (\ref{6}) as
$$
{\cal L} f = -(p-1) f^p.
$$
So,
$$
(\Omega, {\cal L} f) = -(p-1) (\Omega,f^p) $$
where $(\cdot,\cdot)$ denotes
the scalar product in
$L^2 ({\Bbb R}^n,d\mu)$.
By the self-adjointness of ${\cal L}$ and the definition of $\Omega$, i.e.
${\cal L}\Omega = -E\Omega$, we have $$
-E = -(p-1) \frac{(\Omega,f^p)}{(\Omega,f)} < 0 $$
since $\Omega$ and $f$ are strictly positive.

Finally, to see that $f_{\gamma_p} \in {\cal D} ({\cal L})$ we use the fact
that, since $f_{\gamma_p}$ decays at
infinity like (\ref{8}), we have, for any $\delta > 0$, and any $n\in \Bbb N$,
\be
|\nabla_\xi^n
f_{\gamma_p} (\xi) | \leq C(\delta, n) e^{- (\frac{1}{4} - \delta) \xi^2}.
\label{49}
\en
This follows easily from (\ref{8}) and the differential equation (\ref{6}),
and implies that $f_{\gamma_p} \in {\cal D} ({\cal
L}_0)= {\cal D} ({\cal
L})$.\hfill $\Box$

Notice that functions in $L^2 ({\bf
R}^n, d\mu)$ have essentially a
Gaussian decay at infinity,
which is much faster than what is allowed in our Banach space $B$, see
(\ref{19}).
This brings us to the other crucial ingredient in the proof of Proposition
2, which is that $e^{\tau{\cal L}_0}$ contracts functions in $B$ {\it
pointwise} for $\xi$ large. This follows from the explicit formula
(Mehler's formula \cite{Si}):
\be
(e^{\tau{\cal L}_0}) (\xi,\xi') = (4 \pi (1-e^{-\tau}))^{-\frac{n}{2}}
e^{\tau \left( \frac{1}{p-1}-
\frac{n}{2}\right)}
\exp \left(-
\frac{|\xi-e^{-\tau/2}\xi'|^2}{4(1-e^{-\tau})} \right) \label{69}
\en
Hence, if a function $v$ satisfies
\be
|v(\xi)| \leq C(1+|\xi|^q)^{-1},
\label{13}
\en
for some constant $C$, we have
\be
|(e^{\tau {\cal L}_0} v) (\xi)| \leq C' e^{\frac{\tau}{p-1}} (1+|\xi|^q e^{
\frac{\tau q}{2}})^{-1}
\label{14}
\en
for $|\xi|$ large enough (of order $\sqrt \tau$) and another constant $C'$.
Hence, the operator $e^{\tau {\cal L}_0}$ contracts, for large $|\xi|$ and
large $\tau$, any function that decays as in (\ref{13}) with
$q>\frac{2}{p-1}$. By (\ref{fk}), we see that ${\cal L}$
behaves similarly.

The idea of the proof of Proposition 2 is the following. For $|\xi|$ small
(\ref{14}) seems to expand by $e^{\tau\over p-1}$: the potential $V$ is
important in this region and we want to use the information we obtained in
the Hilbert space, Lemma 1 (recall that these functions have rapid decay,
so that this bound should be used to capture the
contraction only in the small $\xi$ region). For large $\xi$ , we shall use
(20). This small-large $\xi$ interplay is however slightly subtle, and we
need to resort to an inductive argument to control the large $\tau$
behaviour in Proposition 2 (this is actually just the Renormalization Group
idea applied to the linear problem).

\vs{3mm}

\no{\bf Proof of Proposition 2}. It is convenient to introduce the
characteristic functions
$$
\chi_s = \chi (|\xi| \leq \rho)
$$
$$
\chi_\ell = \chi (|\xi| > \rho)
$$
where $\rho$ will be chosen suitably below. The properties of ${\cal L}$
that we need are summarized
in the following

\vs{3mm}

\no {\bf Lemma 2}. {\it There exist
constants $C<\infty$, $E>0$, and
$\delta>0$, such that
\begin{enumerate}
\item[i)] For $g \in B$,
\be
\| e^{\tau \cal L} g \|_\infty \leq
C e^{\frac{\tau}{p-1}} \| g \|_\infty.
\label{33}
\en
\item[ii)]
For $g \in L^2 (\Bbb R^n,d \mu)$,
\be
\| e^{\cal L}g \|_\infty \leq C \|g\|_2, \label{35}
\en
where $\| \cdot\|_2 $ is the norm in $L^2 ({\Bbb R}^n, d\mu)$.
\item[iii)]
For $g$ such that $\chi_s g \in L^2 (\Bbb R^n, d \mu)$, \be
\| \chi_\ell e^{\rho \cal L} \chi_s g \|_\infty \leq e^{- \frac{\rho^2}{5}}
\| \chi_s g \|_2,
\label{36}
\en
for $\rho$ large enough.
\item[iv)]
For $g \in B$,
\be
\| \chi_\ell e^{\rho \cal L} g \|_\infty \leq e^{- \delta \rho} \| g
\|_\infty,
\label{37}
\en
for $\rho$ large enough.
\end{enumerate}
}

\vs{3mm}

Let $\|g\|_\infty =1$.
Given Lemma 2, we set $\tau_n=n\rho$, and prove inductively that
there exists $\alpha > 0$
such that $v(\tau_n)=e^{\tau_n{\cal L}}g$ satisfies, for $\rho$ large,
\be
\| \chi_s v (\tau_n) \|_2 + \| \chi_s v(\tau_n)\|_\infty \leq
e^{\frac{\rho^2}{6}} e^{- \alpha n},
\label{39}
\en
and
\be
\| \chi_\ell v (\tau_n) \|_\infty \leq e^{- \alpha n}. \label{40}
\en

Proposition 2 follows from (\ref{39},\ref{40}), by taking
$\mu=\frac{\alpha}{\rho}$ (for times not of the form $\tau=n\rho$, use
(\ref{33})).
The bounds (\ref{39},\ref{40}) hold for $n=0$, for $\rho$ large enough,
using $\|g\|_\infty =1$ and the obvious
inequality
\be
\| \chi_s g \|_2 \leq \rho^{\frac{1}{2}} e^{\frac{\rho^2}{8}} \|
g\|_\infty.
\label{41}
\en

So, let us assume (\ref{39}, \ref{40})
for some $n \geq 0$ and prove it for $n+1$. Let $v = v(\tau_n)$ and write
$$
v = \chi_s v + \chi_\ell v \equiv v_s + v_\ell. $$
We have from Lemma 1 and (25)
\be
\| e^{\rho {\cal L}} v_s \|_2 \leq e^{- \rho E} e^{\frac{\rho^2}{6}} e^{-
\alpha n}
\label{42}
\en
and, combining (\ref{35}) and Lemma 1
\be
\| e^{\rho \cal L} v_s \|_\infty \leq C \| e^{(\rho-1)\cal L} v_s \|_2
\leq C e^{- (\rho-1) E}
e^{\frac{\rho^2}{6}} e^{-\alpha n}.
\label{43}
\en
Finally, from (\ref{33}) and (\ref{40}), we have \be
\| e^{\rho \cal L} v_\ell \|_\infty \leq C e^{\frac{\rho}{p-1}}
e^{-\alpha n}
\label{44}
\en
and, from this and (\ref{41}), we get
\be
\| \chi_s e^{\rho \cal L} v_\ell \|_2 \leq C \rho^{\frac{1}{2}}
e^{\frac{\rho^2}{8}}
e^{\frac{\rho}{p-1}} e^{- \alpha n}.
\label{45}
\en

Combining (\ref{42}-\ref{45}),
one gets (\ref{39}), with $n$ replaced by $n+1$ for $\rho$ large enough and
$\alpha$ small.
On the other hand, (\ref{40}), with $n$
replaced by $n+1$, follows immediately
from (\ref{39},\ref{40}) and (\ref{36},\ref{37}), taking
$\alpha<\delta\rho$.\hfill $\Box$

\vs{3mm}

We are left with the

\vs{3mm}

\no {\bf Proof of Lemma 2}. Part
(i) follows immediately from (\ref{fk}) and (\ref{14}).

For (ii), we use Schwartz' inequality and the bound \be
\sup_\xi \int | e^{\cal L} (\xi , \xi') |^2 d \xi' < \infty, \label{47}
\en
which follows from (\ref{fk}) and (\ref{69}).

For (iii) proceed as in (ii) by using Schwartz' inequality,
but replace
(\ref{47}) by
\be
\sup_{|\xi| > \rho} \left( \int | e^{\rho{\cal L}} (\xi,\xi')|^2 \chi
(|\xi'| \leq \rho) d \xi' \right)^{\frac{1}{2}} \leq e^{- \frac{\rho^2}{5}}
\label{48}
\en
which again follows from (\ref{fk}) and (\ref{69}) (we can replace
$\frac{1}{5}$ in (\ref{48}) by $\frac{1}{4} - \delta$ for any $\delta >0$,
if $\rho$ is large enough).

Finally, (iv) follows from (\ref{14}) and $q>\frac{2}{p-1}$. \hfill $\Box$

\vs{3mm}

\no{\bf Proof of Proposition 1}. The proof is straightforward, given
Proposition 2. We consider the integral equation corresponding to
(\ref{123}):
\be
v(\tau) = e^{\tau{\cal L}} h + \int^\tau_{0} ds e^{(\tau-s){\cal L}} N
(v(s))\equiv {\cal S}(v,\tau) \label{30}
\en
with $v(\tau) \equiv v(\cdot,\tau)$.

First, let $\|h\|_\infty\leq\epsilon$;
(\ref{30}) is solved by the contraction mapping principle in the Banach
space ${\cal B}$ of
functions $v=v(\xi,\tau)$, where $v(\cdot,\tau) \in B$ for $\tau \in
[0,\infty)$, and where the norm is
$$
\| v \| = \sup_{\tau \in [0,\infty)}
\|v (\cdot,\tau) \|_\infty e^{\tau\mu}
$$

We show that $\cal S$
defined by (\ref{30}) maps the ball
$
{\cal B}_0 = \{ v \in {\cal B} | \;
\| v \| \leq C
\varepsilon \}
$
into itself, for a suitable constant $C$, and is a contraction there . This
follows, since we get, using (\ref{123}),
$$
|N(v)| \leq C'|v|^{\tilde p}
$$
where $\tilde p = \min (2,p) >1$, and $C'$ is a constant; therefore
\be
\|N(v(s))\|_\infty\leq C' \|v(s)\|_\infty^{\tilde p} \leq C'
\epsilon^{\tilde p}e^{-{\tilde p}s\mu} \label{38}
\en
and so, Proposition 2 gives the claim for $\varepsilon$ small. The proof
that
${\cal S}$ is a contraction is similar.
Finally,
from the Feynman-Kac formula and the smoothness and boundedness of $V$ one
deduces that
$e^{\tau{\cal L}}(\xi,\xi')=e^{\tau{\cal L}_0}(\xi,\xi') K(\xi,\xi')$ where
$K$ is smooth and bounded. The regularity of the kernel of $e^{\tau{\cal
L}}$ (for short times, it behaves like the heat kernel),
implies that the solution of (34) is
actually the unique classical
solution of (\ref{123}) (for details of such arguments, see \cite{bk3}).

\vs{3mm}

Let now $h\geq 0$.
It is standard that equation (\ref{4}) has unique solution for all times,
with positive initial data such as ours \cite{F}. So,
equation (\ref{123}) also has a unique solution for all times. Moreover, by
a comparison inequality
($v=0$ is a solution of (\ref{123})),
we have $v(\xi, \tau)\geq 0$ for all times. Thus, in (\ref{30}), the second
term is negative ((\ref{123}) shows that $N(v) \leq 0$, and (\ref{fk}, 18)
show that the
kernel of $e^{\tau{\cal L}}$ is positive). Therefore, we have the pointwise
inequality $$
v(\xi, \tau)\leq (e^{\tau{\cal L}}v(0))(\xi) $$
and Proposition 2 proves the claim.\hfill $\Box$

\section{ Extensions and concluding remarks}

1. In Theorems 1 and 2, we could use a norm defined by \be
\| g \|_\infty = {\rm ess}\sup_\xi |g(\xi) w(\xi)|. \en
where $w$ is any bounded positive function decaying at infinity faster than
$|\xi|^{-\frac{2}{p-1}}$. Indeed, the norm (7) was used only in the proof
of (20). However, we would not necessarily get exponential decay of $v$ as
a function of $\tau$.

2. In \cite{Ga} and also in \cite{KP2},
results similar to ours were obtained
on the stability of the self-similar solutions for $p<1+ \frac{2}{n}$ and
$\gamma=\gamma_p$. However, our results apply to
a ball in a Banach space (and to any $\gamma$) while in \cite{Ga} the
initial data is assumed to
satisfy a pointwise inequality (but not to be small). This is similar to
the $h\geq 0$ case in our theorem, but with another inequality.
For the first part of our Theorem, $u(x,1)$, given by (8), does not even
have to be positive.
On the other hand, very general
results on the stability of the self-similar solutions, decaying as in (5),
were obtained in \cite{KP1}, for $p>1+ \frac{2}{n}$. Basically, it
is shown there that any positive initial data decaying at infinity like
$|\xi|^{-
\frac{2}{p-1}}$ will give rise to a solution whose asymptotic behaviour in
time is given by $f_\gamma(\xi)$.

3. With little extra work, the small $\| v \|$ part of the Theorem
generalizes to more general nonlinearities, e.g. equations of the form
\be
\partial_t u = \Delta u - |u|^{p-1} u + F (u, \nabla u) \label{4a}
\en
whereby we need to add to (\ref{123}) the term \be
\tilde F_\tau (v, \nabla v) = e^{\frac{p \tau}{p-1}} F \left(e^{-
\frac{\tau}{p-1}} (f_\gamma +v), e^{- \frac{p+1}{2(p-1)} \tau} \nabla
(f_\gamma +v)\right).
\en

In order to define "small" initial data, we introduce
the Banach
space of $C^1$ functions with the norm:
\be
\| h \| = \| h \|_\infty +
\max_{1 \leq i \leq n} \| \partial_i h \|_\infty \label{20}
\en
where $\| h \|_\infty$ is defined in (7).

We assume that $F$ in (\ref{4a}) is
$C^1$ and
satisfies: \be
|F(a,b)| + |a \partial_a F (a,b)] +
\max_{1 \leq i \leq n} | b_i
\partial_{b_i} F (a,b)| \leq \lambda |a|^{q_1}
\left( \max_{1 \leq i
\leq n} |b_i| \right)^{q_2} \label{21}
\en
for $|a|, |b_i| \leq 1$, where
\be
q_1 + \frac{p+1}{2} q_2 > p
\label{22}
\en
and $\lambda$ is taken small. Using (40, 41), and the fact that $f_\gamma$,
$\partial_i f_\gamma$, belong to the space $B$, one gets
$$
\|\tilde F_s (v, \nabla v)\|_\infty
\leq C\lambda e^{-\delta s}
$$
for some $\delta >0$,
which is like the RHS of (35) for $\lambda$ small. The last two terms in
the LHS of (40) are used to prove that the $\tilde F$ term defines a
contraction. The only extra difficulty is to show that the solution of the
fixed point problem (corresponding to (34)) is a classical solution.
However, using the regularity of the kernel of $e^{\tau{\cal L}}$ discussed
in the proof of Proposition 1, one shows first that $v$ is $C^{2-\alpha}$
for any $\alpha>0$
(its derivative is H\"older
continuous of exponent $1-\alpha$); then, one uses that information, the
integral equation and the regularity of the kernel to show that $v$ is
actually $C^2$. See \cite{bk3} for details.

4. Finally we want to end with some comments on the RG picture behind these
results. In \cite{BKL} the RG map ${\cal R}_L$, for $L>1$ was defined in a
suitable Banach space of initial data $f(x)=u(x,1)$ and a suitable space of
nonlinearities $F$ (taken to be holomorphic functions in \cite{BKL}).
${\cal R}_L$ consisted simply of solving (\ref{2}) up to the finite time
$L^2$ and performing a scale transformation on the solution and on $F$:
\be
{\cal R}_L(f,F)=(f_L,F_L)
\en
where $f_L(x)=L^n u(Lx,L^2)$ and $F_L(u,v,w)=L^{2+n}F(L^{-n}u,
L^{-n-1}v,L^{-n-2}w)$. This scaling assures the semigroup property ${\cal
R}_{L^k}={\cal R}_L^k$ on a common domain. The limit (\ref{1}) was then
shown to follow from \be
{\cal R}_{L}^k (f,F)\rightarrow (Af^*,0) \label{limit}
\en
as $k\rightarrow\infty$, where $(Af^*,0)$ is a one parameter family of
Gaussian ($f^*(x)=e^{-{x^2\over 4}}$) fixed points for ${\cal R}_L$.
Universality, i.e. independence on
$f$ and $F$ was then explained in terms of a dynamical systems picture: if
$(f,F)$ lie on the stable manifold of the line of fixed points, all the
corresponding equations and data have the same asymptotics.

The Theorem of Section 1 is a statement on the stability of a family of
{\it non-Gaussian} fixed points of the RG (which are, moreover,
non perturbative, i.e. we do not use any ``$\varepsilon$-expansion").
Equation (1) is invariant
under the scaling $u \rightarrow u_L$ with $$
u_L (x,t) = L^{\frac{2}{p-1}} u (Lx,L^2t) $$
which suggests setting
$$
f_L (x) = L^{\frac{2}{p-1}} u(Lx,L^2)
$$
and a correspondig definition of $F_L$. Then ${\cal
R}_{L}(f_\gamma,F^*)=(f_\gamma,F^*)$ for $F^*(u)=-u^p$, and our Theorem
constructs the stable manifold of this
fixed point (in the $f$ variable; for $F$, see the previous remark). The
reason one needs an iterative approach to the limit (\ref{limit}), i.e. one
controls the iteration of ${\cal R}_L$ rather than ${\cal R}_L^k$ directly,
is the existence of the neutral direction: $A$ is a nontrivial function of
the data and of the equation. Here there is no neutral direction (this is
the content of Lemma 1), and no iteration is needed
(although we needed to resort to one in
the proof of Proposition 2! But that
was for a different purpose, namely the analysis of the linear operator
$e^{\tau{\cal L}}$).

Combining our results with those of \cite{BKL}, we obtain the following
$p$-dependence of the asymptotics of the solution of $$
\partial_t u = \partial^2 u - u^p,
$$
for a suitable class of initial data. For $p > 1 + \frac{2}{n}$,
$$
u(\cdot
t^{1/2},t) \simeq At^{- \frac{n}{2}} f^* (\cdot) $$
where
$f^*(\xi)=e^{-\xi^2/4}$ is independent of $p$; the prefactor $A$ depends on
$p$ and on the
initial data. We have
$$\int |u(x,t)|dx= {\cal O}(1)$$
as $t \rightarrow \infty.
$
For $p = 1 + \frac{2}{n}$,
$$
u(\cdot t^{1/2},t)
\simeq (A t \log t)^{- \frac{1}{2}}
f^*(\cdot),
$$
i.e. it is as before, but with a
logarithmic
correction and
$$
\int |u(x,t)|dx = {\cal O} ((\log t)^{- 1/2}). $$
For $1<p<1 + \frac{2}{n}$,
$$
u(\cdot t^{1/2}, t)
\simeq t^{- \frac{1}{p-1}} f_\gamma
(\cdot)
$$
where $f_\gamma$ is non-Gaussian,
varies with $p$ but is independent of
the initial data (satisfying the hypotheses of the Theorem).
Moreover,
$$
\int |u(x,t)|dx = {\cal O} \left( t
^{\frac{n}{2}-\frac{1}{p-1}} \right) \rightarrow 0 $$
as $t\rightarrow \infty$.

\vs{3mm}

\no {\bf Acknowledgments}

\vs{3mm}

It is a pleasure to thank J. Miekisz and U. Moschella for interesting
discussions. During this work, J.
Bricmont benefited from the hospitality of the University of Helsinki and
A.
Kupiainen from the one of the University of Louvain. This work was
supported by NSF Grant DMS-8903041 and
by EC Grant SC1-CT91-0695.

\end{document}